\documentclass[manuscript]{aastex}
\newcommand{\lsim}{\, \lower2truept\hbox{${< \atop\hbox{\raise4truept\hbox{$\sim$}}}$}\,}
\newcommand{\gsim}{\, \lower2truept\hbox{${> \atop\hbox{\raise4truept\hbox{$\sim$}}}$}\,}

\newcommand{\oneskip}{\vskip\baselineskip}

\newcommand{\puncspace}{\ifmmode\,\else{\ifcat.\C{\if.\C\else\if,\C\else\if?\C\else%
\if:\C\else\if;\C\else\if-\C\else\if)\C\else\if/\C\else\if]\C\else\if'\C%
\else\space\fi\fi\fi\fi\fi\fi\fi\fi\fi\fi}%
\else\if\empty\C\else\if\space\C\else\space\fi\fi\fi}\fi}
\newcommand{\SP}{\let\\=\empty\futurelet\C\puncspace}

\begin{document}

\title{NIR Luminosity Function of Galaxies in Close Major-Merger Pairs
          and Mass Dependence of Merger Rate}
\author{C.K. Xu}
\affil{Infrared Processing and Analysis Center, 
California Institute of Technology, MS 100-22, Pasadena, CA 91125, USA.
cxu@ipac.caltech.edu}
\author{Y.C. Sun, X.T. He}
\affil{Department of Astronomy, Beijing Normal University,
Beijing 100875, China}
\received{2003 October 10}
\accepted{2004 January 28}

\begin{abstract}
A sample of close major-merger pairs 
(projected separation ${\rm 5 \leq r \leq 20 h^{-1}}$ kpc, ${\rm K_s}$
band magnitude difference $\delta {\rm K_s} \leq 1$ mag)
is selected from the
matched 2MASS-2dFGRS catalog of Cole et al. (2001). The pair primaries
are brighter than ${\rm K_s} = 12.5$ mag. After corrections for 
various biases, the comparison between counts in the paired galaxy
sample and counts in the parent sample shows that for 
the local `M$_*$ galaxies' sampled by flux limited surveys, 
the fraction of galaxies in the close major-merger pairs is 1.70$\pm
0.32\%$.  Using 38 paired
galaxies in the sample, a ${\rm K_s}$ band luminosity function (LF) is
calculated. This is the first unbiased LF for a sample of objectively 
defined interacting/merging galaxies in the local universe,
while all previously determined LFs of paired galaxies are
biased by mistreating paired galaxies as singles.
A stellar mass function (MF) is translated from the
LF. Compared to the LF/MF of 2MASS galaxies, a differential pair
fraction function is derived. The results suggest a trend in
the sense that less massive galaxies may have
lower chance to be involved in close major-merger pairs than more
massive galaxies. The algorithm presented in this paper
can be easily applied to much larger samples of 2MASS galaxies with
redshifts in near future.
\end{abstract}


\section{Introduction}

Galaxy-galaxy interactions/mergers are
very important in the cosmic evolution of galaxies. In the
hierarchical galaxy formation paradigm, galaxies and galactic
structures are formed through merging of smaller galaxies/structures
(Kauffmann et al. 1993; Cole et al. 2000).  There is strong evidence
that galaxy-galaxy interactions/mergers can significantly enhance the
star formation rate (SFR) in galaxies involved (see Kennicutt 1998 for a
review).  There is also evidence that the strong evolution in the
cosmic SFR is due to a population of peculiar/interacting starburst
galaxies which are closely related to galaxy mergers (Brinchmann et
al. 1998). Theoretical
simulations (Barnes 1990) and observations (Schweizer 1978;
Kormendy \& Sanders 1992) show
that gas-rich late-type galaxies transform to gas-poor
early-type E/S0 galaxies through galaxy mergers. This process may also
be responsible for the formation of bulges of disk galaxies. 
The AGN activity, another physical process  
linking the mass to the light, may also be
closely related to galaxy-galaxy interactions/mergers if
the central black holes in AGNs are built up mostly in
galaxy mergers, given the tight correlation between the black hole
mass and the bulge mass (Franceschini et al. 1999).
During the merging, the tidal torque can
send large amount of gas and stars deep into the galactic nuclear
region, and therefore feed a pre-existing black hole very efficiently,
leading to enhanced AGN activity. In summary,
galaxy-galaxy interactions/mergers play a central role in four of the most
important processes in galaxy evolution, including mass assembly, star
formation, morphological transformation and AGN activity.

There are two classical methods to select interacting/merging
galaxies. One is to identify binary galaxies. And the other is to find
galaxies with peculiar morphology (e.g. with tidal tails, double
nuclei, or distorted disc). The latter has the advantage of including
both pre- and post-mergers, and the identifications need not to be
confirmed by redshift data. Studies based on this method found strong
evolution in the fraction of major-mergers 
(galaxy pairs with mass ratio $\lsim 3$), particularly for massive
galaxies (Brinchmann et al. 1998; Le F\'evre et
al. 2000; Conselice et al. 2003). However, these results may suffer
uncertainties due to difficulties in quantifying the morphological
peculiarity and in detecting the interaction signatures in 
high z galaxies because of poor spatial resolutions and the cosmic dimming. 
In contrast, it is easy to define binary galaxies quantitatively and
objectively. This makes an objectively defined comparison between
local merger events and high z merger events possible. However,
earlier studies of pair fraction and its cosmic evolution have
suffered seriously from the contamination of unphysical pairs because
of the lack of redshifts or highly incomplete redshift data (Zepf \&
Koo 1989; Burkey et al 1994; Carlberg et al. 1994; Yee \& Ellington
1995; Woods et al. 1995; Patton et al. 1997; Wu \& Keel 1998). 
In two recent studies using samples of galaxies with
measured redshifts, Le F\'evre et al (2000) and Patton et al. (2002)
found $m=2.7\pm 0.6$ and $m=2.3\pm 0.7$, respectively, where m is the
evolution index in the evolution function of the merger rate: $R_{mg}
\propto (1+z)^m$.  In a series of
papers, Patton et al. (1997; 2000; 2002) pointed out that in studies
of merger rate and evolution, it is very important to correct various
systematic biases, and results from comparisons between
mismatched samples of low-z and high-z galaxies are not very
meaningful. 

The best way to constrain the merger rate and its cosmic evolution
is to compare differential pair fraction functions at different redshifts. 
A differential pair fraction function (DPFF) is defined by ratios 
between number of paired galaxies and
the total number of galaxies in luminosity (or stellar mass)
bins. Such functions are not sensitive to sample selection
(flux limited or volume limited) and therefore can be compared
without bias between different studies.
DPFF can be determined by comparing the luminosity (mass) function
of paired galaxies with that of total galaxies.  In this paper, we 
estimate the local ${\rm K_s}$ (2.16$\mu m$) band luminosity function of
close major-merger pairs, and derive from it the
DPFF in the z=0 universe. The very close relation between
the ${\rm K_s}$ band luminosity and the stellar mass means that for
the first time we can have the mass function of the paired galaxies
and the mass dependence of the merger rate. Since this can be compared
directly to the predictions of the hierarchical galaxy formation
simulations (e.g. Benson et al. 2002), it will provide an important
test for these simulations. Through out this paper, 
we adopt the $\Lambda$-cosmology 
with $\Omega_m=0.3$ and $\Omega_\Lambda = 0.7$, 
$h = {\rm H}_0/(100\; {\rm km~sec}^{-1} {\rm Mpc}^{-1})$.

\section{Sample Selection}
The parent sample is selected from the matched 2MASS-2dFGRS catalog of
Cole et al. (2001), which has 45289 galaxies with measured
J (1.25$\mu m$), H(1.65$\mu m$) and ${\rm K_s}$ (2.16$\mu m$) magnitudes from
the Extend Source Catalog (XSC) of 2MASS. The default 
${\rm K_{20}}$ magnitude is used for the ${\rm K_s}$ band
fluxes (Jarrett et al. 2000). Among 45289 galaxies, 17173 have measured
redshifts from the 2dFGRS survey. 
It is known that the coverage of the 2dFGRS
survey is not uniform, and there are holes between individual
2dF fields (Colless et al 2001). In order to minimize the
uncertainties in our pair statistics due to the uneven redshift
coverage, we restrict the parent sample to galaxies which have the
redshift completeness index $c_{z} \geq 0.5$, where $c_z$ is defined
as the ratio between the number of galaxies with measured redshifts
and the total number of galaxies within 1 deg radius from the center
of the galaxy
in question. Our final parent sample has 19053 galaxies, of which 14083 
have measured redshifts (74\% redshift completeness). The number
counts of all galaxies and of galaxies with
measured redshifts in the parent sample are plotted in
Fig.1. Apparently the sample is complete down to ${\rm K_s}$ = 13.5, which is
the completeness limit of the XSC
(Jarrett et al. 2000). There is no significant dependence of the 
redshift completeness on the ${\rm K_s}$ magnitude.

In the pair selection procedure, we search for neighbors around every
galaxy with a measured redshift in the parent sample. 
Neighbors are not required to have measured redshifts.
Among the matches, we
select pairs according to the following criteria:
(1) The ${\rm K_s}$ magnitude of the primary is not fainter than 12.5.
A primary is defined as the brighter component of a pair. 
(2) At least one of the components has measured redshift.
(3) When both components have measured redshifts, 
the velocity difference is not larger than 500 km sec$^{-1}$.
(4) The projected separation is in the range of 
${\rm 5 \leq r \leq 20 h^{-1}}$ kpc.
When only one component has measured redshift, the separation 
is calculated according to that redshift and the angular distance 
between the components.
(5) The ${\rm K_s}$ difference between the two components
is not larger than 1 magnitude. (6) Not in clusters.
According to criteria (1) and (5), all selected galaxies 
in the pair sample 
are brighter than ${\rm K_s}$=13.5, the completeness limit of the parent
sample. This ensures the completeness of the pair sample. Criteria (4)
and (5) restrict our pairs to ``close major-merger pairs". This not
only reduces the contamination of unphysical pairs among those with
only one measured redshift, but also confines our sample to pairs
which have high probability to merge within a few $10^8$ years (Patton
et al. 2000). Our final sample has 19 galaxy pairs (38 galaxies). Among
them, 3 pairs have both components with measured redshifts, and 16 have
only one component with measured redshift. The redshift range of the
galaxy pairs is $0.016 < z <0.070$ with a median of z=0.039.

\section{Pair Fraction and ${\rm K_s}$ Band Luminosity Function}

Two biases in our pair 
selection have to be corrected. First, pairs of both components
without measured redshift are missed in our sample. 
Given the minimum fiber separation of the redshift
survey ($\sim 30''$, Colless et al. 2001)
and the median separation of our pairs ($21''$),
it is very likely that the fraction of
missing pairs is significantly higher than 
estimate from Poisson statistics (7\%). 
An empirical approach is taken to estimate this fraction:
in the parent sample there are 350 pairs of galaxies with
projected separation $< 30''$, of which
31.7\%  having both components without measured redshift. 
Secondly, the 16 single-redshift pairs are contaminated by unphysical
pairs. In order to estimate how many false pairs are expected, Monte
Carlo simulations reproducing the number counts and the redshift
distribution of the parent sample are carried out. 
Coordinates in a 650 deg$^2$ sky region are randomly assigned to the
simulated sources. A pair selection procedure, including all
criteria listed in Section 2 except for (3), 
is applied to the simulated samples. A total of 100 such
simulations are carried out. From this we found a mean of 6.36
with a standard dispersion of 2.62 for the total number of
unphysical pairs. Since we miss 31.7\% of all pairs, and
one such false pair (with both components having measured 
redshift, not included in the pair
sample) is already found 
in the real case, it is expected that $3.34=6.36\times (1-0.317) -1$
(with an error of $1.79=2.62\times (1-0.317)$) unphysical pairs are
among the 16 single-redshift pairs. Furthermore, the chance of
being a false pair is proportional to the searching area,
which is inversely proportional to $z^2$, multiplied by $n$
which is the local density ($r \leq 10'$) 
of neighboring galaxies of $\rm |K_s -K'_s| \leq 1$, 
with $\rm K'_s$ being the magnitude of the seed galaxy. 
This is reflected in the following `false factor':
\begin{equation}
Q_{false,i} = \left\{ \begin{array}{ll}
               0 \pm 0\;\; & 
               \mbox{(pairs with 2 redshifts)} \\
               (3.34 \pm 1.79) \times 
               (n_i/z_i^2)/ \sum_{j} (n_j/z_j^2) \;\; & 
               \mbox{(pairs with 1 redshift),}
              \end{array}
\right.
\end{equation}
where the summation is over the 16 single-redshift pairs.

The pair fraction can be estimated as follows:
\begin{equation}
f_{p} = {A\over N_{g}}\times \sum_i^{N_{pg}} (1-Q_{false,i})
\end{equation}
where $N_g = 2079$ is the total number of galaxies in the parent
sample brighter than ${\rm K_s}$=12.5; 
$A=1/(1-0.317)$ is the correction factor to compensate
the incompleteness due to the missing pairs;
$N_{pg} = 30$ is the total number
of galaxies in the paired galaxy sample brighter than ${\rm K_s}$=12.5. 
The estimated error of $f_{p}$ is
$err = {A\over N_{g}}\times\sqrt{  \sum_i^{N_{pg}} 
[(1-Q_{false,i})^2 +e_{Q,i}^2]}$,
where $e_Q$ is the error of $Q_{false}$ as given in Eq(1).
Note that the two galaxies in a pair have the same $Q_{false}$ and $e_Q$. 
Using these formulae, we found a pair fraction of $f_p = 1.70\pm 0.32 \%$. 

The ${\rm K_s}$ band luminosity 
function (LF) of paired galaxies is calculated using
the ${\rm V_{max}}$ method (Schmidt 1968). 
Comparing the number counts of our parent
sample with the 2MASS number counts of Kochanek et al. (2001) in
Fig.1, we estimate that the effective sky coverage of the parent
sample is 650 deg$^2$, with an error of $\sim 5\%$.
We will ignore this error because it is much smaller than
other errors. Given our pair selection criteria,
both components of
a pair have the same ${\rm V_{max}}$ determined by the ${\rm K_s}$
magnitude of the primary, the redshift of the pair, and 
${\rm K_{lim}=12.5}$. The ${\rm K_s}$ band luminosity
function and its error are calculated using the following
formulae:
\begin{equation}
{\rm 
\phi(M_{K,i}) = {A \over \delta(m)} 
\sum_{j}^{N_i} {1-Q_{false,j} \over V_{max,j}} ;
}
\end{equation}
\begin{equation}
{\rm
e_{\phi}(M_{K,i}) = {A \over \delta(m)} 
\sqrt{\sum_{j}^{N_i} {(1-Q_{false,j})^2 +e_{Q,j}^2 
\over V_{max,j}^2}} ;
}
\end{equation}
where ${\rm \phi(M_{K,i})}$ is the luminosity 
function in the i-th bin of the ${\rm K_s}$
band absolute magnitude; ${\rm N_i}$ 
is the number of galaxies in that bin; 
${\rm \delta(m)} = 0.5$ is the bin width; 
${\rm V_{max,j}}$ is the 
maximum finding volume of the j-th galaxy in the
bin. Other symbols have the same definitions as in Eq(2) and Eq(3).
The results are listed in Table 1 and plotted in Fig.2. The
parameters of the best-fit Schechter
function of the LF are given
in Table 2. It is well known that LF derived using ${\rm V_{max}}$
method can be affected by inhomogeneous spatial distribution of
galaxies. In the redshift distribution of the 2MASS-2dFGRS matched
catalog (Cole et al. 2001), there is evidence of clustering,
particularly a dip around z=0.04 and a sharp peak around
z=0.06. Therefore, the fluctuations of the LF of paired galaxies
around the smooth Schechter function (e.g. the excess at M$_{\rm
K} = -22.75$) are possibly due to this effect.
   
The stellar masses,
corresponding to the absolute magnitude bins of the LF, are also listed
in Table 1. Following Kochanek et al. (2001) and
Cole et al. (2001), we first translate the isophotal ${\rm K_s}$ magnitude to
the 'total' ${\rm K_s}$ magnitude ($\delta {\rm K_s} =0.2$ mag), 
then assume the conversion factor of ${\rm M_{\rm star}/L_{\rm
K} = 1.32 M_\sun/L_\sun}$ which is for a Salpeter IMF (Cole et
al. 2001).  The differential pair fraction function (Table 1 and Fig.2)
is calculated using the Schechter functions of
the paired galaxies and of 2MASS galaxies
(Kochanek 2001) in the luminosity/mass range
covered by the paired galaxy sample, with the error estimated from
the quadratic sum of the error of
the LF of paired galaxies and its deviation from
 the Schechter function. The last bin is not included
because there is only a single galaxy in the bin and therefore
the value is too uncertain. Although error bars are
substantial, some general trends can be
identified in Fig.2. Unlike the LF of 2MASS galaxies, 
the LF of paired galaxies has a
negative slope in the faint end, suggesting that galaxies with low
stellar mass are less likely to be involved in the close major-merger
pairs. Indeed the DPFF shows a significant trend in
the sense that in the low mass bins the pair fraction is low.
This result appears to be in contradiction with the
simple 'chance hypothesis', which states that if pairs
are formed by single galaxies falling into each other's
gravitational influence zone through random motion, then
the fraction of major merger pairs among less massive galaxies should
be relatively high because less massive galaxies are more abundant
than more massive galaxies. The major reason for relatively low
pair fractions in the low mass bins is perhaps  
the gravitational perturbations due to massive neighbors.

\section{Comparison with Earlier Studies}

Many authors have attempted to estimate the pair fraction in the local
universe (Zepf et al. 1989; Burkey et al. 1994; Carlberg et al. 1994;
Yee et al. 1994; Patton et al. 1997; Patton et al. 2000). Patton et
al. (2000), using the SSRS2 sample of galaxies with redshifts, derived
a pair fraction of 2.26$\pm 0.52\%$ for close pairs.  
This is slightly higher than
what is found in this work (${\rm f_p}=1.70 \pm 0.32\%$).  However, we have
restricted our pair sample to major-merger pairs ($\delta {\rm K_s} \leq 1$,
corresponding to a mass ratio of $ \leq 2.5$). In the pair sample of
Patton et al. (2000), there is no restriction for the magnitude
difference while an absolute magnitude range
of $\rm -21 < M_B < -18$ is imposed. Since missing secondaries is perhaps the
most important source of incompleteness of a pair sample, we
strongly argue that a maximum magnitude difference
should be applied in any analysis on pair statistics. The restriction
to major-merger pairs also makes the comparison to results of studies
on peculiar galaxies easier because those galaxies are mostly
major-mergers (Conselice et al. 2003).

Xu \& Sulentic (1991), Rabaca \& Sulentic (1994), Soares et al. (1995)
and Hernandez Toledo et al. (1999)
calculated the B band luminosity
functions of paired galaxies. These results suggest  
a rather constant pair fraction of $\sim 4$ --- 10\%. The
most serious bias in these analyses is due to the
mistreatment of paired galaxies as singles.
Namely, the ${\rm V_{max}}$ of each paired galaxy is
derived from its own B band magnitude, with the cutoff of ${\rm
B_{lim}} = 14.5$. In this case, the
quantity $1/{\rm V_{max}}$ is an {\it incorrect} 
estimator of the contribution of
a paired galaxy to the density of the population. This is because
the finding of one component of a pair depends on the finding of the
other component of the same pair (otherwise neither of the two
galaxies will be included in the paired galaxy sample), and the two
components have different maximum finding volumes due to different
luminosities. Therefore, contrary to the definition of ${\rm V_{max}}$, the
component with the larger ${\rm V_{max,1}}$ may not be found in such a
volume because the other component (with a smaller ${\rm V_{max,2}}$) will
be missing beyond ${\rm V_{max,2} < V_{max,1}}$. Keel \& Wu (1995) derived a
LF for galaxy mergers selected morphologically. 
The weakness of that work is a rather
heterogeneous and arbitrary sample selection.

In summary, the ${\rm K_s}$ band luminosity function of galaxies in
close major-merger pairs presented in this paper is the first unbiased
luminosity function of a sample of objectively defined
interacting/merging galaxies in the local universe.  The algorithm
presented in this paper can be easily applied to much larger samples
of 2MASS galaxies with redshifts in near future when the SDSS redshift
survey data and the 6dF survey data are fully released. The local mass
function of paired galaxies will be compared with those of the high-z
galaxies when deep SIRTF survey data in the MIR bands are available.
The rest frame ${\rm K_s}$ band radiation of galaxies of z=0.6 and z=1
will be detected in the SIRTF-IRAC 3.6$\mu m$ band and 4.5$\mu m$ band,
respectively. Extensive redshift surveys for these galaxies have
been planned (Lonsdale et al. 2003). Comparisons between mass
functions of paired galaxies in the $z=0,\, 0.6$ and 1 universe will
reveal the evolution of merger rate and its mass dependence, and
provide strong constraints to theories of galaxy formation and
evolution.


We thank Roc Cutri for reading and commenting on an earlier version
of this paper.


\clearpage
\noindent{\bf Table 1.} ${\rm K_s}$ Band Luminoisty Function of Paired Galaxies and Differential Pair Fraction

\hskip-1.5truecm\begin{tabular}{ccccccc}
\hline
\hline
(1) & (2) & (3) & (4) & (5) & (6) & (7) \\
${\rm M_K - 5\log(h)}$ & $\log({\rm M_{stars}/h^{-2})}$ 
& $\log(\phi/{\rm h}^{3})$ 
& error & N & pair fraction & error \\
 (mag) & (M$_\sun$) & (${\rm Mpc^{-3} mag^{-1}}$) & & & (\%) & \\
\hline
&&&&&&\\
 -21.75 &  10.24 & -4.25 &   -4.35  &   2   &    0.78  &   0.50 \\
 -22.25 &  10.44 & -4.07 &   -4.31  &   4   &    1.27  &   0.65 \\
 -22.75 &  10.64 & -3.76 &   -4.20  &   9   &    1.89  &   1.25 \\
 -23.25 &  10.84 & -3.94 &   -4.42  &  11   &    2.45  &   0.84 \\
 -23.75 &  11.04 & -4.38 &   -4.74  &   6   &    2.54  &   1.17 \\
 -24.25 &  11.24 & -4.64 &   -4.99  &   5   &    1.85  &   0.97 \\
 -24.75 &  11.44 & -5.60 &   -5.60  &   1   &          &        \\
&&&&&&\\
\hline
\hline
\end{tabular}
\oneskip

\vskip1truecm


\noindent{\bf Table 2.} Paremeters of Schechter Function of ${\rm K_s}$ LF of Paired Galaxies
\nopagebreak

\hskip-0.5truecm\begin{tabular}{cccccc}\hline
\hline
&&&&& \\
 $\alpha$ & error & ${\rm M_* - 5\log(h)}$ & error 
& ${\rm \log(\phi_0/h^{3})}$ & error\\
&&&&& \\
\hline
    0.30  &  0.56 & -22.55 & 0.25 &  -3.46 & 0.13 \\
&&&&& \\
\hline
\hline
\end{tabular}

\clearpage

\begin{figure}
\plotone{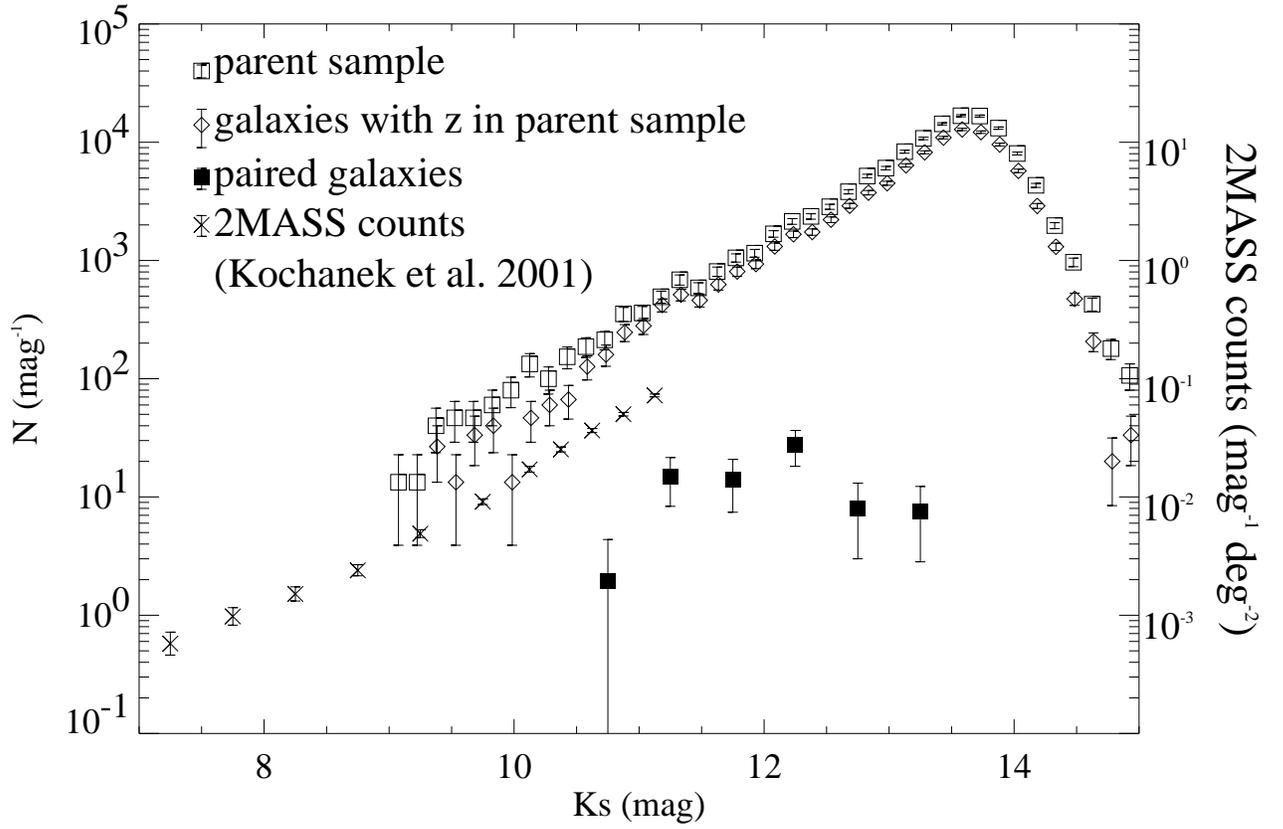}
\caption{${\rm K_s}$ band number counts.}
\label{fig1}
\end{figure}

\begin{figure}
\plotone{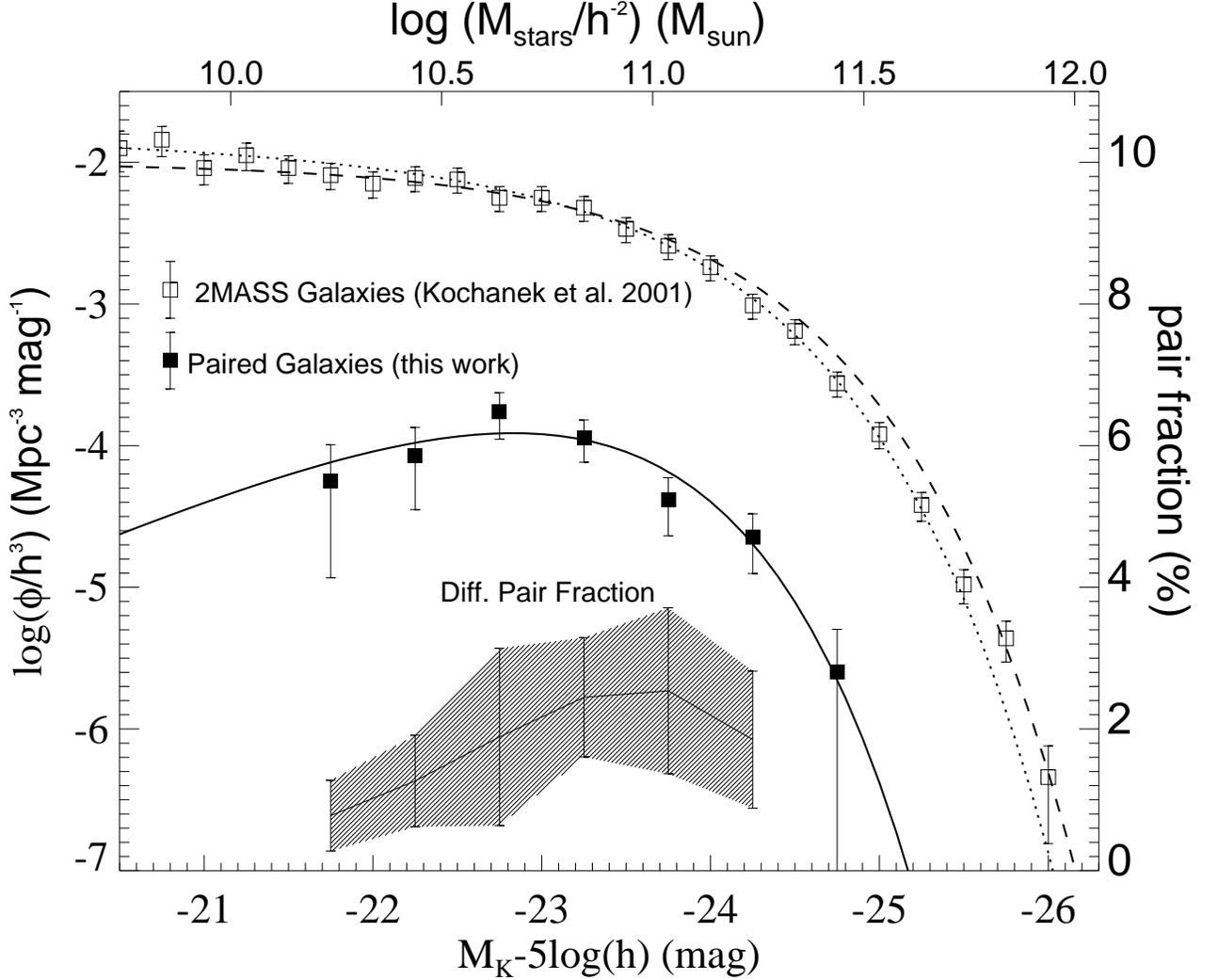}
\caption{${\rm K_s}$ band luminosity functions and stellar mass
functions, and differential pair fraction (right coordinates). The
lines are Schechter functions of the LF of paired galaxies (solid), 
of the LF of 2MASS galaxies by Kochanek et al. (2001) (dotted), 
and of the LF of 2MASS galaxies by Cole et al. (2001) (dashed). 
The shaded area presents the differential pair fractions and
the errors.}
\label{fig2}
\end{figure}
\end{document}